\documentstyle[12pt,a4]{article}
\input epsf

\global\arraycolsep=2pt

\begin{document}

\begin{titlepage}

\begin{flushright}
CERN-TH.7395/94\\
hep-ph/9408290
\end{flushright}

\vspace{0.3cm}

\begin{center}
\Large\bf Theoretical Update on the\\
Model-Independent Determination of $|V_{cb}|$\\
Using Heavy Quark Symmetry
\end{center}

\vspace{0.8cm}

\begin{center}
Matthias Neubert\\
{\sl Theory Division, CERN, CH-1211 Geneva 23, Switzerland}
\end{center}

\vspace{0.8cm}

\begin{abstract}
In view of new precise measurements of the $\bar B\to
D^*\ell\,\bar\nu$ decay rate near zero recoil, we reconsider the
theoretical uncertainties in the extraction of $|\,V_{cb}|$ using
heavy quark symmetry. In particular, we combine our previous

estimate of $1/m_Q^2$ corrections to the normalization of the
hadronic form factor at zero recoil with sum rules derived by Shifman
{\it et al}.\ to obtain a new prediction with less theoretical
uncertainty. We also summarize the status of the calculation of
short-distance corrections, and of the slope of the form factor at
zero recoil. We find ${\cal F}(1)=\eta_A\,\widehat\xi(1)=0.93\pm
0.03$ and $\widehat\varrho^2=0.7\pm 0.2$. Combining this with the
most recent experimental results, we obtain the model-independent
value $|\,V_{cb}|=0.040\pm 0.003$.
\end{abstract}

\vspace{0.8cm}
\centerline{(submitted to Physics Letters B)}

\vspace{1.5cm}
\noindent
CERN-TH.7395/94\\
August 1994

\end{titlepage}

\section{Introduction}

With the discovery of heavy quark symmetry (for a review see
Ref.~\cite{review} and references therein), it has become clear that
the study of exclusive semileptonic $\bar B\to D^*\ell\,\bar\nu$
decays close to zero recoil allows for a reliable determination of
the CKM matrix element $V_{cb}$, which is free, to a large extent, of
hadronic uncertainties \cite{Volo}--\cite{Vcb}. Model dependence
enters this analysis only at the level of power corrections, which
are suppressed by a factor of at least $(\Lambda_{\rm QCD}/m_c)^2$.
These corrections can be investigated in a systematic way using the
heavy quark effective theory \cite{Geor}. They are found to be small,
of the order of a few per cent.

Until recently, this method to determine $|\,V_{cb}|$ was limited by
large experimental uncertainties of about 15--20\%, which were much
larger than the theoretical uncertainties in the analysis of
symmetry-breaking corrections. However, three collaborations have now
presented results of higher precision \cite{CLEO}--\cite{ARGUS}. It
is thus important to reconsider the status of the theoretical
analysis, even more so since the original analysis of power
corrections in Ref.~\cite{FaNe} has become the subject of some
controversy \cite{Shif}.

Besides reviewing some of the existing calculations, the main purpose
of this note is to propose a ``constructive synthesis'' of the two
approaches that have been suggested to obtain an estimate of the
power corrections to the decay form factor at zero recoil. These
corrections are parametrized by a quantity $\delta_{1/m^2}$. The
``exclusive approach'' of Falk and myself \cite{FaNe} has the
advantage that it provides an exact expression for $\delta_{1/m^2}$
involving five hadronic parameters, which are defined in terms of
matrix elements of higher-dimensional operators in the heavy quark
effective theory. The final numerical estimate is model-dependent,
since four of these five parameters are not precisely known. The
``inclusive approach'' of Shifman {\it et al}.\ \cite{Shif} provides
an upper bound for $\delta_{1/m^2}$ in terms of only two parameters;
however, it is not clear to which extent this bound is saturated. We
shall combine the two approaches and derive non-trivial constraints
on the hadronic parameters in the formula for $\delta_{1/m^2}$. These
constraints help to reduce the theoretical uncertainty.

Let us start with a short discussion of the decay kinematics
\cite{review}. The hadronic matrix element describing the decay
process $\bar B\to D^*\ell\,\bar\nu$ can be parametrized by invariant
helicity amplitudes corresponding to transverse and longitudinal
polarization of the $D^*$ meson. As kinematic variable, we choose the
product of the meson velocities, $w=v_B\!\cdot\!v_{D^*}$, which is
related to the momentum transfer $q^2$ to the lepton pair by
\begin{equation}
   w = {m_B^2 + m_{D^*}^2 - q^2\over 2 m_B m_{D^*}} \,.
\end{equation}
The differential decay rate ${\rm d}\Gamma/{\rm d}w$ is proportional
to the sum over the squared helicity amplitudes, which up to a
kinematic factor defines the square of a function ${\cal F}(w)$. The
resulting expression is
\begin{eqnarray}\label{BDrate}
   {{\rm d}\Gamma\over{\rm d}w}
   &=& {G_F^2\over 48\pi^3}\,(m_B-m_{D^*})^2\,m_{D^*}^3
    \sqrt{w^2-1}\,(w+1)^2 \nonumber\\
   &&\times \bigg[ 1 + {4w\over w+1}\,
    {m_B^2-2w\,m_B m_{D^*} + m_{D^*}^2\over(m_B - m_{D^*})^2}
    \bigg]\,|\,V_{cb}|^2\,{\cal F}^2(w) \,.
\end{eqnarray}
The heavy quark effective theory allows the factorization of the
short- and long-distance contributions to ${\cal F}(w)$ into a
perturbative coefficient $\eta_A$ and a hadronic form factor
$\widehat\xi(w)$:
\begin{equation}
   {\cal F}(w) =\eta_A\,\widehat\xi(w) \,.
\end{equation}
In the heavy quark limit, this hadronic form factor agrees with the
Isgur--Wise function $\xi(w)$ \cite{Isgu,Falk}. We use the notation
$\widehat\xi(w)$ to indicate that the two functions differ by terms
suppressed by inverse powers of the heavy quark masses. Luke's
theorem determines the normalization of $\widehat\xi(w)$ at zero
recoil ($w=1$) up to second-order power corrections \cite{Luke}:
\begin{equation}
   \widehat\xi(1) = 1 + \delta_{1/m^2} \,.
\end{equation}
The strategy proposed in Ref.~\cite{Vcb} is to measure the product
$|\,V_{cb}|\,{\cal F}(w)$ as a function of $w$, and to extrapolate it
to $w=1$ to extract
\begin{equation}
   |\,V_{cb}|\,{\cal F}(1) = |\,V_{cb}|\,\eta_A\,
   (1 + \delta_{1/m^2}) = |\,V_{cb}|\,\Big\{ 1
   + O\big[ \alpha_s(m_Q),1/m_Q^2 \big] \Big\} \,,
\end{equation}
where we use $m_Q$ as a generic notation for $m_c$ or $m_b$. The task
of theorists is to provide a reliable calculation of the
symmetry-breaking corrections contained in $\eta_A$ and
$\delta_{1/m^2}$ in order to turn this measurement into a precise
determination of $|\,V_{cb}|$. In Sect.~\ref{sec:2}, we briefly
review the status of the calculation of short-distance corrections. A
new theoretical analysis of power corrections is given in
Sect.~\ref{sec:3}. In Sect.~\ref{sec:4}, we give a theoretical
prediction for the slope of the form factor $\widehat\xi(w)$ at zero
recoil. This parameter is important for the extrapolation of
experimental data to $w=1$. Section~\ref{sec:5} contains a summary of
the numerical results and some conclusions.

\section{Calculation of $\eta_A$}
\label{sec:2}

The short-distance coefficient $\eta_A$ takes into account a finite
renormalization of the axial vector current in the region
$m_b>\mu>m_c$. Its calculation is a straightforward application of
QCD perturbation theory. At the one-loop order, one finds
\cite{Volo,QCD1,Pasc}
\begin{equation}
   \eta_A = 1 + {\alpha_s\over\pi}\,\bigg(
   {m_b + m_c\over m_b - m_c}\,\ln{m_b\over m_c} - {8\over 3}
   \bigg) \,.
\end{equation}
The scale of the running coupling constant is not determined at this
order. Choosing $\alpha_s$ between $\alpha_s(m_b)\simeq 0.20$ and
$\alpha_s(m_c)\simeq 0.32$, and using $m_c/m_b=0.30\pm 0.05$, one
obtains values in the range $0.95<\eta_A<0.98$. The scale ambiguity
leads to an uncertainty of order $\Delta\eta_A\sim[(\alpha_s/\pi)
\ln(m_b/m_c)]^2\sim 2\%$.

The calculation can be improved by using the renormalization group to
resum the leading and next-to-leading logarithms of the type
$[\alpha_s\ln(m_b/m_c)]^n$, $\alpha_s [\alpha_s\ln(m_b/m_c)]^n$, and
$(m_c/m_b) [\alpha_s\ln(m_b/m_c)]^n$ to all orders in perturbation
theory \cite{PoWi}--\cite{FaGr}. A consistent scheme for a
next-to-leading-order calculation of $\eta_{\em A}$ has been
developed in Ref.~\cite{QCD2}. The result is
\begin{eqnarray}\label{etaA}
   \eta_A = x^{6/25}\,\bigg\{ 1
    &+& 1.561\,{\alpha_s(m_c)-\alpha_s(m_b)\over\pi}
    - {8\alpha_s(m_c)\over 3\pi} \nonumber\\
   &+& {m_c\over m_b}\,\bigg( {25\over 54}
    - {14\over 27}\,x^{-9/25} + {1\over 18}\,x^{-12/25}
    + {8\over 25}\,\ln x \bigg) \nonumber\\
   &+& {2\alpha_s(m)\over\pi}\,
    {m_c^2\over m_b(m_b - m_c)}\,\ln{m_b\over m_c} \bigg\} \,,
\end{eqnarray}
where $x=\alpha_s(m_c)/\alpha_s(m_b)$, and $m_b>m>m_c$. The numerical
result is very stable under changes of the input parameters. Using
$\Lambda_{\overline{\rm MS}}=(0.25\pm 0.05)$ GeV (for four flavours)
and $m_c/m_b=0.30\pm 0.05$, one obtains $\eta_A=0.985\pm 0.006$. The
uncertainty arising from next-to-next-to-leading corrections is of
order $\Delta\eta_A\sim(\alpha_s/\pi)^2\sim 1\%$.

Equation (\ref{etaA}) is an exact result to a given order in
perturbation theory. We stress that, since next-to-leading effects
are properly included, it is not only valid for large values of
$\ln(m_b/m_c)$. Therefore, we disagree with the criticism of this
calculation by the authors of Ref.~\cite{Shif}. Of course, it would
be desirable to know the non-logarithmic terms of order $\alpha_s^2$
in $\eta_A$, but we see no reason why these terms should be unusually
large. Taking this usual perturbative uncertainty into account, we
believe it is conservative to increase the error by a factor 2.5 and
quote
\begin{equation}
   \eta_A = 0.985\pm 0.015 \,.
\end{equation}

\section{Anatomy of $\delta_{1/m^2}$}
\label{sec:3}

Hadronic uncertainties enter the determination of $|\,V_{cb}|$ at the
level of second-order power corrections, which are expected to be of
order $(\Lambda_{\rm QCD}/m_c)^2\sim 3\%$. For a precision
measurement of $|\,V_{cb}|$, it is important to understand the
structure of these corrections in detail. In our discussion (as in
all previous analyses), we will investigate the $1/m_Q^2$ corrections
at the tree level, thereby neglecting effects of order
$\alpha_s(m_Q)/m_Q^2$. In particular, we will not discuss the running
of the hadronic parameters of the effective theory. In view of the
theoretical uncertainty in the estimate of these non-perturbative
parameters, this is a safe approximation.

Using the technology of the heavy quark effective theory, Falk and
myself have derived the exact expression \cite{FaNe}
\begin{equation}\label{delm2}
   \delta_{1/m^2} = - \bigg( {1\over 2 m_c} - {1\over 2 m_b}
    \bigg) \bigg( {\ell_V\over 2 m_c} - {\ell_P\over 2 m_b}
    \bigg) + {1\over 4 m_c m_b}\,\bigg( {4\over 3}\,\lambda_1
    + 2\lambda_2 - \lambda_{G^2} \bigg) \,,
\end{equation}
which depends upon five hadronic parameters that are independent of
the heavy quark masses. They have the following physical
significance: $\ell_P$ and $\ell_V$ parametrize the deficit in the
``wave-function overlap'' between $b$- and $c$-flavoured pseudoscalar
({\it P\/}) and vector ({\it V\/}) mesons. For instance, $\ell_P$ is
defined as
\begin{equation}
   \langle D(v)|\,c^\dagger b\,|\bar B(v)\rangle = 2 v^0\,\eta_V\,
   \bigg\{ 1 - \bigg( {1\over 2 m_c} - {1\over 2 m_b} \bigg)^2\,
   \ell_P + O(1/m_Q^3) \bigg\} \,,
\end{equation}
where $\eta_V\simeq 1.03$ is a short-distance correction factor
\cite{review}, and we use a mass-independent normalization of meson
states. The corresponding relation for vector mesons defines
$\ell_V$. The parameter $\lambda_1=-\langle\vec p_Q^{\,2}\rangle$ is
proportional to the kinetic energy of the heavy quark inside a heavy
meson, and $\lambda_2={1\over 4}(m_V^2-m_P^2)$ determines the
vector--pseudoscalar mass splitting arising from operators in the
effective Lagrangian that break the heavy quark spin symmetry. From
the observed mass splitting between $B$ and $B^*$ mesons, one obtains
$\lambda_2 \simeq 0.12$ GeV$^2$. Finally, $\lambda_{G^2}$
parametrizes certain matrix elements containing two insertions of
operators that break the spin symmetry. In our analysis below, we
will assume that this parameter is small, i.e.\ of a magnitude
similar to $\lambda_2$ or smaller. This assumption is supported by
QCD sum rule calculations of other spin-symmetry-breaking corrections
to heavy quark decay form factors \cite{subl1,subl2}.

With the exception of $\lambda_2$, estimates of these hadronic
parameters are model-dependent. In Ref.~\cite{FaNe}, we made the
simplifying assumptions that $\ell_P=\ell_V$, and that the
corrections represented by $\lambda_{G^2}$ are negligible. Using then
reasonable values such as $\ell_P=\ell_V=(0.35\pm 0.15)$ GeV$^2$ and
$-\lambda_1=(0.25\pm 0.20)$ GeV$^2$, one obtains
$\delta_{1/m^2}=-(2.4\pm 1.3)\%$. Here and in the following, we take
$m_b=4.80$ GeV and $m_c=1.45$ GeV for the heavy quark masses. In
Ref.~\cite{review}, the error in the estimate of $\delta_{1/m^2}$ has
been increased to $\pm 4\%$ in order to account for the model
dependence and higher-order corrections. A very similar result,
$-5\%<\delta_{1/m^2}<0$, has been obtained by Mannel \cite{Mann}.

Recently, Shifman {\it et al}.\ have suggested an alternative
approach to obtain an estimate of $\delta_{1/m^2}$ \cite{Shif}. The
idea is to apply an operator product expansion to the $\bar B$-meson
matrix element of the time-ordered product of two flavour-changing
heavy quark currents, and to equate the resulting theoretical
expression to a phenomenological expression obtained by saturating
the matrix element with physical intermediate states. This leads to
sum rules, which can be used to derive inequalities for the $\bar
B\to D^{(*)}$ transition form factors at zero recoil. In
Ref.~\cite{Shif}, such bounds have been obtained for the parameters
$\ell_P$ and $\delta_{1/m^2}$. They are
\begin{eqnarray}\label{sumrul1}
   \ell_P &>& \ell_P^{\rm min} > 0 \,, \nonumber\\
   \\
   \delta_{1/m^2} &<& - \bigg( {1\over 2 m_c} - {1\over 2 m_b}
    \bigg) \bigg( {\ell_V^{\rm min}\over 2 m_c}
    - {\ell_P^{\rm min}\over 2 m_b} \bigg)
    + {1\over 4 m_c m_b}\,\bigg( {4\over 3}\,\lambda_1
    + 2\lambda_2 \bigg) \nonumber\\
   &<& - {\lambda_2\over 2 m_c^2}\simeq -2.9\% \,, \nonumber
\end{eqnarray}
where
\begin{equation}
   \ell_P^{\rm min} = {1\over 2}(-\lambda_1 - 3\lambda_2) \,,
   \qquad \ell_V^{\rm min} = {1\over 2}(-\lambda_1 + \lambda_2) \,.
\end{equation}
The first relation in (\ref{sumrul1}) implies that \cite{Vol}
\begin{equation}\label{lam1bou}
   -\lambda_1 > 3\lambda_2\simeq 0.36~\mbox{GeV}^2 \,,
\end{equation}
excluding some of the values for the parameter $\lambda_1$ used in
previous analyses of $\delta_{1/m^2}$. It implies that the average
heavy quark momentum inside the heavy meson is quite large, of order
600 MeV. Ball and Braun have calculated $\lambda_1$ using QCD sum
rules and find $-\lambda_1=(0.5\pm 0.1)$ GeV$^2$ \cite{BaBr}. Below
we shall use $\lambda_1=-0.4$ GeV$^2$. We will comment on the (weak)
dependence of our results on the value of $\lambda_1$ later. The
upper bound for $\delta_{1/m^2}$ in (\ref{sumrul1}) implies
that\footnote{In Ref.~\protect\cite{Shif}, this number is quoted as
0.94.} $\eta_A\,\widehat\xi(1) < 0.956$. Of course, a crucial
question is to what extent this inequality is saturated. The authors
of Ref.~\cite{Shif} make an ``educated guess'' that $\eta_A\,
\widehat\xi(1)=0.89\pm 0.03$ corresponding to $\delta_{1/m^2}=
-(9.6\pm 3.0)\%$. However, the arguments presented to support this
guess are not very rigorous.

\begin{figure}
\epsfxsize=8cm
\centerline{\epsffile{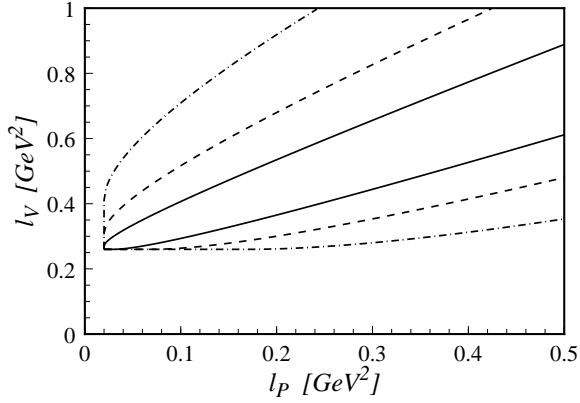}}
\centerline{\parbox{13cm}{\caption{\label{fig:1}
Allowed regions in the $\ell_P$--$\ell_V$ plane for
$\lambda_{G^2}=0.01$ GeV$^2$ (solid), 0.05 GeV$^2$ (dashed), and 0.15
GeV$^2$ (dash-dotted). We use $\lambda_1=-0.4$~GeV$^2$, in which
case $\ell_P^{\rm min}=0.02$ GeV$^2$ and $\ell_V^{\rm min}=0.26$
GeV$^2$.}}}
\end{figure}

It seems more appealing to us to use the sum rules to constrain the
hadronic parameters in (\ref{delm2}). We first note that it is
possible to derive two additional relations by interchanging
pseudoscalar with vector meson states, corresponding to transitions
of the type $\bar B^*\to D^{(*)}$. Repeating the derivations of
Ref.~\cite{Shif} for this case, we find
\begin{eqnarray}\label{sumrul2}
   \ell_V &>& \ell_V^{\rm min} > 2\lambda_2 \simeq
    0.24~\mbox{GeV}^2 \,, \nonumber\\
   \\
   \widetilde\delta_{1/m^2} &<& - \bigg( {1\over 2 m_c}
    - {1\over 2 m_b} \bigg) \bigg( {\ell_P^{\rm min}\over 2 m_c}
    - {\ell_V^{\rm min}\over 2 m_b} \bigg)
    + {1\over 4 m_c m_b}\,\bigg( {4\over 3}\,\lambda_1
    + 2\lambda_2 \bigg) \,, \nonumber
\end{eqnarray}
where $\widetilde\delta_{1/m^2}$ is obtained from $\delta_{1/m^2}$ in
(\ref{delm2}) by interchanging $\ell_P$ with $\ell_V$. The first
relation in (\ref{sumrul2}) puts a bound on the parameter $\ell_V$.
To obtain further constraints, we use the fact that the above
relations are valid for an arbitrary value of the mass ratio
$m_c/m_b$. Comparing the second relation in (\ref{sumrul1}) with
(\ref{delm2}) in the limit $m_b=m_c$, we find that
\begin{equation}
   \lambda_{G^2} > 0 \,.
\end{equation}
Moreover, as long as $m_c<m_b$, it follows that
\begin{eqnarray}
   \ell_V - \ell_V^{\rm min} &>& {m_c\over m_b}\,
    (\ell_P - \ell_P^{\rm min}) - {m_c\over m_b - m_c}\,
    \lambda_{G^2} \,, \nonumber\\
   \ell_V - \ell_V^{\rm min} &<& {m_b\over m_c}\,
    (\ell_P - \ell_P^{\rm min}) + {m_b\over m_b - m_c}\,
    \lambda_{G^2} \,.
\end{eqnarray}
We are free to choose any value of the mass ratio $m_c/m_b$ between 0
and 1 to make these relations as restrictive as possible. It is then
straightforward to show that
\begin{equation}\label{compl}
   \mbox{max}\Big\{ \sqrt{\ell_P-\ell_P^{\rm min}}
   - \sqrt{\lambda_{G^2}} \,;\,\, 0 \Big\} <
   \sqrt{\ell_V-\ell_V^{\rm min}} <
   \sqrt{\ell_P-\ell_P^{\rm min}} + \sqrt{\lambda_{G^2}} \,.
\end{equation}
For small values of $\lambda_{G^2}$, this relation implies a
correlation between $\ell_P$ and $\ell_V$, which is such that
$\ell_V-\ell_P \simeq \ell_V^{\rm min}-\ell_P^{\rm min} = 2\lambda_2
\simeq 0.24$ GeV$^2$. This is illustrated in Fig.~\ref{fig:1}, where
we show the allowed regions in the $\ell_P$--$\ell_V$ plane for
different values of $\lambda_{G^2}$. In total, we have thus
identified three effects, which decrease $\delta_{1/m^2}$ with
respect to the estimate given in Ref.~\cite{FaNe}: a large value of
$(-\lambda_1)$, a positive value of $\lambda_{G^2}$, and the fact
that $\ell_V$ is likely to be larger than $\ell_P$ provided that
$\lambda_{G^2}$ is small.

\begin{figure}
\epsfxsize=8cm
\centerline{\epsffile{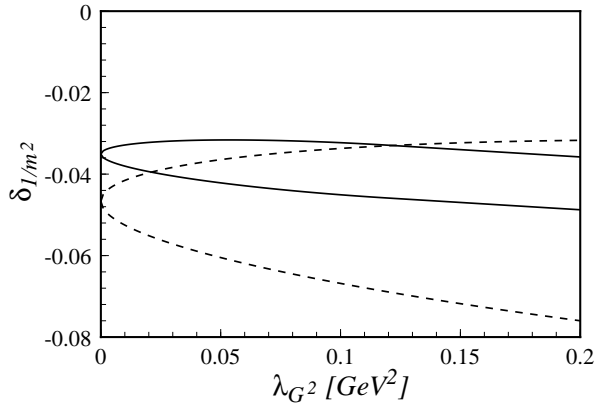}}
\centerline{\parbox{13cm}{\caption{\label{fig:2}
Allowed regions for $\delta_{1/m^2}$ as a function of $\lambda_{G^2}$
for the two cases $\bar\ell=0.2$ GeV$^2$ (solid) and 0.4 GeV$^2$
(dashed).}}}
\end{figure}

To proceed, it is convenient to introduce new parameters
\begin{eqnarray}
   \bar\ell &=& {1\over 2}\,(\ell_V + \ell_P) \,, \nonumber\\
   S &=& {1\over 2}\,\Big\{ (\ell_V - \ell_V^{\rm min})
    + (\ell_P - \ell_P^{\rm min}) \Big\}
    = \bar\ell + {1\over 2}\,(\lambda_1+\lambda_2) \,, \\
   D &=& {1\over 2}\,\Big\{ (\ell_V - \ell_V^{\rm min})
    - (\ell_P - \ell_P^{\rm min}) \Big\}
    = {1\over 2}\,(\ell_V - \ell_P) - \lambda_2 \,, \nonumber
\end{eqnarray}
in terms of which
\begin{eqnarray}\label{final}
   \delta_{1/m^2} &=& - \bigg( {1\over 2 m_c} - {1\over 2 m_b}
    \bigg)^2\,\bar\ell - \bigg( {1\over 4 m_c^2} - {1\over 4 m_b^2}
    \bigg) (\lambda_2 + D) \nonumber\\
   &&\mbox{}+ {1\over 4 m_c m_b}\,\bigg( {4\over 3}\,\lambda_1
    + 2\lambda_2 - \lambda_{G^2} \bigg) \,.
\end{eqnarray}
The fact that $S>0$ implies $\bar\ell > {1\over 2}(-\lambda_1
-\lambda_2)>\lambda_2$. The inequality (\ref{compl}) is equivalent
to $-D_{\rm max} < D < D_{\rm max}$, where
\begin{equation}
   D_{\rm max} = \left\{ \begin{array}{cl}
    S & ;\quad 0 < S\le\lambda_{G^2}/2 \,, \\
    \sqrt{\lambda_{G^2} S - \lambda_{G^2}^2/4} &
    ;\quad S\ge\lambda_{G^2}/2 \,.
   \end{array} \right.
\end{equation}
The main uncertainty in evaluating (\ref{final}) comes from the
unknown values of the parameters $\bar\ell$ and $\lambda_{G^2}$. As a
guideline, one may employ the constituent quark model of Isgur {\it
et al}.\ \cite{ISGW}, in which one uses non-relativistic harmonic
oscillator wave functions for the ground-state heavy mesons, for
instance $\psi_B(r)\sim \exp(-{1\over 2}\mu\omega r^2)$, where
$\mu=(1/m_q+1/m_b)^{-1}$ is the reduced mass. One then obtains
$\bar\ell={3\over 4} m_q^2\simeq 0.2$ GeV$^2$, where we take
$m_q\simeq 0.5$ GeV for the light constituent quark mass,
corresponding to the difference between the spin-averaged meson
masses and the heavy quark masses. However, this estimate of
$\bar\ell$ is probably somewhat too low. Lattice studies of
heavy-light wave functions suggest an exponential behaviour of the
form $\psi_B(r)\sim\exp(-\kappa\mu r)$ \cite{latt}, which leads to
$\bar\ell={3\over 2} m_q^2\simeq 0.4$ GeV$^2$. We believe that values
much larger than this are unlikely, since we use a rather large
constituent quark mass $m_q$. In fact, adopting the point of view
that the sum rules for $\ell_P$ and $\ell_V$ are saturated to
approximately 50\% by the ground-state contribution \cite{Shif}, one
would expect $\bar\ell\simeq (-\lambda_1-\lambda_2)\simeq 0.28$
GeV$^2$, which seems a very reasonable value to us. In
Fig.~\ref{fig:2}, we show the allowed regions for $\delta_{1/m^2}$ as
a function of $\lambda_{G^2}$ for two values of $\bar\ell$. When
$-\lambda_1$ is varied between 0.36 and 0.5 GeV$^2$, the resulting
values for $\delta_{1/m^2}$ change by less than 1\%. For all
reasonable choices of parameters, the results are in the range
$-8\%<\delta_{1/m^2}<-3\%$. Hence, we quote our new value as
\begin{equation}
   \delta_{1/m^2} = - (5.5\pm 2.5)\% \,,
\end{equation}
which is consistent with the previous estimates in
Refs.~\cite{FaNe,Shif,Mann} at the $1\sigma$ level. A more precise
determination of the parameter $\bar\ell$ would help to reduce the
uncertainty in this number.

We conclude this section with a word of caution. Recently, it has
been shown \cite{Chris} that the sum rules derived by Shifman {\it et
al}.\ in Ref.~\cite{Shif} suffer from a renormalon ambiguity; in
other words, they do not obey the renormalization-group equation if
the theory is regulated with a hard momentum cutoff. This is a
serious problem, which has to be solved before these sum rules can be
used with confidence in phenomenological applications. Here, we
assume that the renormalon problem can be cured without changing the
form of the sum rules.

\section{Prediction for the slope parameter $\widehat\varrho^2$}
\label{sec:4}

In the extrapolation of the differential decay rate (\ref{BDrate})
to zero recoil, the slope of the function $\widehat\xi(w)$ close to
$w=1$ plays an important role. One defines a slope parameter
$\widehat\varrho^2$ by
\begin{equation}\label{rhohat}
   \widehat\xi(w) = \widehat\xi(1)\,\Big\{ 1
   - \widehat\varrho^2\,(w-1) + O\big[ (w-1)^2 \big] \Big\} \,.
\end{equation}
It is important to distinguish $\widehat\varrho^2$ from the
corresponding slope parameter $\varrho^2$ of the Isgur--Wise
function. They differ by corrections that break the heavy quark
symmetry. Whereas the slope of the Isgur--Wise function is a
universal, mass-independent parameter, the slope of the physical form
factor depends on logarithms and inverse powers of the heavy quark
masses. On the other hand, $\widehat\varrho^2$ is an observable
quantity, while the value of $\varrho^2$ depends on the
renormalization scheme.

To establish the relation between the two parameters, it is
convenient to introduce in an intermediate step the axial vector form
factor $h_{A_1}(w)$ defined as
\begin{equation}
   \langle D^*(v_{D^*},\epsilon)|\,\bar c\,\gamma^\mu\gamma_5 b\,
   |\bar B(v_B)\rangle = (w+1)\,h_{A_1}(w)\,\epsilon^{*\mu}
   + \ldots \,,
\end{equation}
where the ellipses represent terms proportional to $v_B^\mu$ or
$v_{D^*}^\mu$. The relation between the physical form factors
$\widehat\xi(w)$ and $h_{A_1}(w)$ is given in Ref.~\cite{review}.
Defining a slope parameter $\varrho_{A_1}^2$ in analogy to
(\ref{rhohat}), we find
\begin{equation}\label{delrho}
   \Delta\varrho^2 = \widehat\varrho^2 - \varrho_{A_1}^2
   = - {1\over 6}\,(R_1^2-1) - {1\over 3}\,
   {m_B\over m_B - m_{D^*}}\,(1-R_2) \,,
\end{equation}
where $R_1$ and $R_2$ denote certain ratios of the $\bar B\to D^*$
decay form factors at zero recoil \cite{subl1}. In the heavy quark
limit, $R_1=R_2=1$, and the two slope parameters coincide. The
symmetry-breaking corrections to these ratios have been analysed in
detail. Including both short-distance and $1/m_Q$ corrections, which
in this case can be calculated without much model dependence, one
obtains $R_1=1.3\pm 0.1$ and $R_2=0.8\pm 0.1$ \cite{review}. From
(\ref{delrho}), it then follows that $\Delta\varrho^2=-(0.22\pm
0.06)$. Recently, the form factor ratios $R_1$ and $R_2$ have been
measured by the CLEO collaboration, with the result that $R_1=1.30\pm
0.39$ and $R_2=0.64\pm 0.29$ \cite{Kuts}. This leads to
$\Delta\varrho^2=-(0.31\pm 0.20)$, in nice agreement with our
theoretical prediction.

The next step is to relate the form factor $h_{A_1}(w)$ to the
Isgur--Wise function. The matrix element that defines the Isgur--Wise
function in the heavy quark effective theory is ultraviolet-divergent
(for $w\ne 1$) and needs to be regularized by introducing a
subtraction scale $\mu$. To leading order in $1/m_Q$, the regularized
function $\xi(w,\mu)$ is related to the physical form factor
$h_{A_1}(w)$ by a renormalization-group-invariant Wilson coefficient
function $\widehat C_1^5$, which contains the dependence on the heavy
quark masses, and a universal function $K_{\rm hh}$ containing the
dependence on the renormalization scale \cite{QCD2}:
\begin{equation}
   h_{A_1}(w) = \widehat C_1^5(m_b,m_c,w)\,K_{\rm hh}(w,\mu)\,
   \xi(w,\mu) + O(1/m_Q) \,.
\end{equation}
These functions are known to next-to-leading order in
renormalization-group-improved perturbation theory. Using the
explicit expression for $K_{\rm hh}(w,\mu)$ given in
Ref.~\cite{review}, we find that the physical slope parameter
$\varrho_{A_1}^2$ is related to the slope parameter $\varrho^2(\mu)$
of the regularized Isgur--Wise function by
\begin{eqnarray}
   \varrho_{A_1}^2 &=& \varrho^2(\mu) + {16\over 81}\,
    \ln\alpha_s(\mu) + {8\over 81}\,\bigg( {94\over 9} - \pi^2
    \bigg)\,{\alpha_s(\mu)\over\pi} \nonumber\\
   &&\mbox{}- \bigg[ {\partial\over\partial w}
    \ln\widehat C_1^5(m_b,m_c,w) \bigg]_{w=1} + O(1/m_Q) \nonumber\\
   &\equiv& \varrho^2 - \bigg[ {\partial\over\partial w}
    \ln\widehat C_1^5(m_b,m_c,w) \bigg]_{w=1} + O(1/m_Q) \,,
\end{eqnarray}
where the last equation defines the $\mu$-independent slope
$\varrho^2$ of the renormalized Isgur--Wise function at
next-to-leading order. Using the explicit expression for the Wilson
coefficient given in Ref.~\cite{QCD2}, one finds that
$\varrho_{A_1}^2=\varrho^2 + (0.21\pm 0.02) + O(1/m_Q)$. An estimate
of the $1/m_Q$ corrections to this relation is model-dependent. We
shall not attempt it, but instead increase the theoretical
uncertainty to $\pm 0.2$. Hence, we obtain
\begin{equation}\label{rhorel}
   \widehat\varrho^2 = \varrho_{A_1}^2 - (0.22\pm 0.06)
   \simeq \varrho^2 \pm 0.2 \,.
\end{equation}

Theoretical predictions for the renormalized slope parameter
$\varrho^2$ have been obtained from QCD sum rules, including a
next-to-leading-order renormalization-group improvement. These
calculations are tedious, since it is necessary to include two-loop
radiative corrections to resolve the issue of scheme dependence. The
complete calculation of these corrections has been performed in
Ref.~\cite{twoloop}. It leads to $\varrho^2=0.7\pm 0.1$
\cite{review}. A similar result has been found by Bagan {\it et al}.\
\cite{Baga}. Based on (\ref{rhorel}), we thus predict
\begin{equation}
   \widehat\varrho^2 = 0.7\pm 0.2 \,.
\end{equation}

\section{Summary}
\label{sec:5}

The exclusive semileptonic decay mode $\bar B\to D^*\ell\,\bar\nu$
provides for the cleanest determination of the CKM matrix element
$V_{cb}$. Heavy quark symmetry can be used to calculate the
differential decay rate close to zero recoil in a model-independent
way, up to small symmetry-breaking corrections, which can be analysed
in a systematic expansion in powers of $\alpha_s(m_Q)$ and $1/m_Q$
using the heavy quark effective theory. In this note, we have
reconsidered and updated the analysis of these corrections. We find
$\eta_A=0.985\pm 0.015$ for the Wilson coefficient of the axial
vector current, and $\delta_{1/m^2} = -(5.5\pm 2.5)\%$ for the power
corrections to the normalization of the function $\widehat\xi(w)$ at
zero recoil. The latter value is new and has been obtained by
combining the existing approaches to estimate these corrections in a
constructive way. Using these results, we predict
\begin{equation}\label{etaxi}
   {\cal F}(1) = \eta_A\,\widehat\xi(1) = 0.93\pm 0.03
\end{equation}
for the normalization of the hadronic form factor ${\cal F}(w)$ at
zero recoil.

Three experiments have recently presented new measurements of the
product $|\,V_{cb}|\,{\cal F}(1)$. When rescaled using the new
lifetime values $\tau_{B^0}=(1.61\pm 0.08)$ ps and
$\tau_{B^+}=(1.65\pm 0.07)$~ps \cite{Roud}, the results obtained from
a linear fit to the data are\footnote{The ARGUS result has also been
corrected for the new $D$ branching fractions \protect\cite{Ritch}.}
\begin{equation}
   |\,V_{cb}|\,\eta_A\,\widehat\xi(1) = \left\{
   \begin{array}{ll}
   0.0347\pm 0.0019\pm 0.0020 &
    ;\quad \mbox{CLEO~\protect\cite{CLEO},} \\
   0.0382\pm 0.0044\pm 0.0035 &
    ;\quad \mbox{ALEPH~\protect\cite{ALEPH},} \\
   0.0388\pm 0.0043\pm 0.0025 &
    ;\quad \mbox{ARGUS~\protect\cite{ARGUS},}
   \end{array} \right.
\end{equation}
where the first error is statistical and the second systematic.
Following the suggestion of Ref.~\cite{Ritch}, we add $0.001\pm
0.001$ to these values to account for the curvature of the function
$\widehat\xi(w)$. Using then the theoretical result (\ref{etaxi}), we
obtain
\begin{equation}\label{Vcbval}
   |\,V_{cb}| = 0.0399\pm 0.0026\,(\mbox{exp})
   \pm 0.0013\,(\mbox{th}) = 0.0399\pm 0.0029 \,,
\end{equation}
which corresponds to a model-independent measurement of $|\,V_{cb}|$
with 7\% accuracy. This is by far the most accurate determination to
date.

We disagree with the conclusion of Ref.~\cite{Shif} that inclusive
$b\to c\,\ell\,\bar\nu$ decays would allow for a more reliable
determination of $|\,V_{cb}|$. In this case, one has to make an
assumption about the heavy quark masses that appear in the
theoretical expression for the inclusive decay rate even at leading
order. Moreover, it has been demonstrated that the extraction of
$|\,V_{cb}|$ from inclusive decays suffers from a perturbative
uncertainty of about 10\%, due to unknown higher-order corrections in
the expansion in $\alpha_s(m_Q)$ \cite{BaNi}. Nevertheless, the most
recent values obtained from the analysis of $b\to c\,\ell\,\bar\nu$
decays, which are \cite{Ritch}
\begin{equation}
   |\,V_{cb}| = \left\{
   \begin{array}{ll}
   0.039\pm 0.001\,(\mbox{exp})\pm 0.005\,(\mbox{th}) &
    ;\quad \mbox{measurements at $\Upsilon(4s)$,} \\
   0.042\pm 0.002\,(\mbox{exp})\pm 0.005\,(\mbox{th}) &
    ;\quad \mbox{measurements at $Z^0$,}
   \end{array} \right.
\end{equation}
are in excellent agreement with (\ref{Vcbval}). The theoretical
uncertainty in these numbers is larger than in the extraction from
exclusive decays, however, and it is harder to control.

Finally, we have related the physical slope parameter
$\widehat\varrho^2$ to the slope of the Isgur--Wise function and
obtain the prediction $\widehat\varrho^2=0.7\pm 0.2$ based on
existing QCD sum rule calculations. It is consistent with the average
value observed by experiments, which is $\widehat\varrho^2=0.87\pm
0.12$ \cite{CLEO}--\cite{ARGUS}.

\subsection*{Acknowledgements}

It is a pleasure to thank R.~Patterson for providing me with the most
recent experimental numbers on $|\,V_{cb}|$, and T.~Mannel for useful
discussions.


\begin{thebibliography}{99}
\bibitem {Vol}
This bound was first pointed out to me by M. Voloshin.
\end{thebibliography}
\end{document}